\pdfoutput=1
\documentclass[aps,prc,reprint,floatfix,longbibliography,nofootinbib,groupedaddress,showkeys]{revtex4-1}

\usepackage{amsmath}    
\usepackage{graphicx}   
\usepackage{color}
\usepackage{siunitx}
\usepackage[colorlinks,linkcolor=blue,citecolor=blue, 
urlcolor=blue]{hyperref}   
\usepackage{bm} 
\usepackage{import}

\def\dd{{d}}
\def\x{{\bm x}}
\def\k{{\bm k}}

\def\p{{\bf p}}

\newcommand\nda{\end{align}}
\def\half{{\textstyle\frac{1}{2}}}

\def\st{\begin{equation}}
\def\stp{\end{equation}}
\def\bg{\begin{eqnarray}}
\def\nd{\end{eqnarray}}
\def\Eq#1{Eq.~(\ref{#1})}

\def\Fig#1{Fig.~\ref{#1}}
\def\Sect#1{Sec.~\ref{#1}}

\def\llangle{\left\langle}
\def\rrangle{\right\rangle}

\advance\parskip 1.9pt
\advance\voffset -0.2in

\def\st{\begin{equation}}
\def\stp{\end{equation}}
\def\bg{\begin{eqnarray}}
\def\nd{\end{eqnarray}}

\def\element{\in}
\def\parts{{N_{\rm parts}}}

\begin{document}

\title{Subleading harmonic flows in hydrodynamic simulations of heavy ion collisions}

\author{Aleksas Mazeliauskas}
\email[]{aleksas.mazeliauskas@stonybrook.edu}
\affiliation{Department of Physics and Astronomy, Stony Brook University, New 
York 11794, USA}

\author{Derek Teaney}
\email[]{derek.teaney@stonybrook.edu}
\affiliation{Department of Physics and Astronomy, Stony Brook University, New 
York 11794, USA}

\date{\today}

\begin{abstract}
We perform a principal component analysis (PCA) of $v_3(p_T)$ in
event-by-event hydrodynamic simulations of Pb+Pb collisions at the Large Hadron 
Collider (LHC).  The
PCA procedure identifies two dominant contributions to the two-particle
correlation function, which together capture 99.9\% of the squared variance.
We find that the subleading flow (which is the largest source of flow
factorization breaking in hydrodynamics) is predominantly a response to the
radial excitations of a third-order  eccentricity.  We present a systematic
study of the hydrodynamic response to these radial excitations in 2+1D viscous
hydrodynamics.  Finally, we construct a good geometrical predictor for the
orientation angle  and magnitude of the leading and subleading flows using two
Fourier modes of the initial  geometry.
\end{abstract}


\maketitle


\section{Introduction}

Two-particle correlation measurements in ultrarelativistic  heavy ion
collisions provide an extraordinarily detailed test of the hydrodynamic
description of heavy ion events. Indeed, the measured two-particle correlations
exhibit elliptic, triangular, and higher harmonic flows, which can be used to
constrain the transport properties of the quark gluon plasma (QGP) produced in
heavy ion collisions~\cite{Heinz:2013th,Luzum:2013yya}.  In hydrodynamic
simulations of heavy-ion events, fluctuations in the initial state are
propagated by the expansion dynamics of the QGP, and this expansion ultimately
induces fluctuations in the momentum spectra of the produced particles.  Thus,
measurements of  the momentum space fluctuations (or correlations) constrain
the properties of the QGP expansion and the initial state.  The purpose of the
current paper is to classify and quantify the dominant momentum space
fluctuations in (boost-invariant) event-by-event hydrodynamics, and then to
optimally correlate these fluctuations  in momentum space with specific
fluctuations in the initial state geometry.  The current paper is focused on
triangular flow, since it is a strong signal and driven entirely by fluctuations~\cite{Alver:2010gr}. The corresponding studies of the other harmonics are
postponed for future work.

Due to flow fluctuations the correlation matrix of event-by-event triangular
flows, $\llangle V_{3}(p_{T1}) V_3^*(p_{T2}) \rrangle$, in hydrodynamics does not
factorize~\cite{Gardim:2012im}.  Factorization breaking is quantified by the
parameter $r(p_{T1}, p_{T2})$,
\st
\label{rmatrix}
r(p_{T1}, p_{T2}) \equiv \frac{ \llangle V_{3}(p_{T1})  V_{3}^*(p_{T2}) \rrangle}{\sqrt{ \llangle |V_{3}(p_{T1})|^2 \rrangle \llangle |V_{3}(p_{T2})|^2 \rrangle} } \leq 1 \, ,
\stp
which must be less than unity when there are several statistically independent
sources of triangular flow in the event sample~\cite{Gardim:2012im}.
Factorization breaking has been studied in event-by-event
hydrodynamics~\cite{Gardim:2012im,Heinz:2013bua,Kozlov:2014fqa} and compares
reasonably to the measured data for appropriate
parameters~\cite{Kozlov:2014fqa}. It is generally understood from these
analyses that factorization breaking is caused by the hydrodynamic response to
geometrical properties of the initial state that are poorly characterized by
the coarse geometrical measure $\varepsilon_3$.  For instance, in
Ref.~\cite{Kozlov:2014fqa} the $r(p_{T1},p_{T2})$ matrix was found to be sensitive to
a parameter controlling the roughness of the initial state.  In
Ref.~\cite{Heinz:2013bua} it was suggested that a careful study of the $r$ matrix and
other observables could be used to test  hydrodynamic predictions for the
$p_{T}$ dependence of the event plane angle, which arises when multiple
triangular flows are present in a single event.  The current paper clarifies
the origin of factorization breaking by associating the largest
nonfactorizable contribution to the triangular flow with the hydrodynamic
response to the first radial excitation in the triangular  geometry. 

First, in \Sect{PCA_sect} we use principal component analysis (PCA) of the
harmonic spectrum to analyze the transverse momentum dependence of the third
harmonic in boost invariant event-by-event hydrodynamics. PCA is a statistical
technique that decomposes the flow correlation matrix into eigenvectors and
eigenvalues~\cite{Bhalerao:2014mua}. The procedure naturally identifies the
most important contributions to flow fluctuations.  Typically  only two  modes
are needed to give an excellent  description of the full covariance matrix
$\llangle V_3(p_{T1}) V_3^*(p_{T2}) \rrangle$ to 0.1\% accuracy.  When there are
only two significant eigenvectors (or triangular flow patterns), the $r$ matrix
can be expressed as~\cite{Bhalerao:2014mua}
\st
\label{pcar}
r(p_{T1},p_{T2})           \simeq 1 - \frac{1}{2} \left( \frac{V_3^{(2)}(p_{T1})}{V_3^{(1)}(p_{T1}) } -  \frac{V_3^{(2)}(p_{T2})}{V_3^{(1)}(p_{T2})}\right)^2 \, ,
\stp
where  $V_{3}^{(1)}(p_T)$  and $V_{3}^{(2)}(p_T)$ are the 
first and second eigenvectors%
\footnote{
As described in \Sect{PCA_sect}, the eigenvectors are normalized to the
eigenvalue $\int_0^{\infty} \dd p_{T} (V_3^{(a)}(p_T))^2 = \lambda_a$, and we
are assuming that $\lambda_1 \gg \lambda_2 $.}. 
 
The leading mode of the third harmonic is strongly correlated with the triangular event
plane\cite{Alver:2010gr}
, and thus is essentially equivalent to familiar measurements of
$v_3(p_T)$ with the scalar product or event plane method.  However, the
subleading mode is uncorrelated with the leading event plane, and is therefore
projected out in most measurements of harmonic flow.  Section \ref{subleading} 
studies
the basic properties of the subleading triangular flow, such as its dependence
on centrality and viscosity. 

In \Sect{avg_geometry} we show that the subleading triangular flow arises
(predominantly) from the radial excitation of the triangular geometry.  To
reach this conclusion we first directly calculate the average geometry in the
event plane of the leading and subleading flows. This averaged geometry (as
explained in \Sect{avg_geometry}) is shown in \Fig{PCA_S_1}  and exhibits a
familiar triangular shape for the leading flow and a triangular shape with a
radial excitation for the subleading flow. 

Having identified the physical origin of the subleading flow, we  introduce
several geometric predictors which (with various degrees of accuracy)
quantitatively predict the magnitude and orientation of the subleading flow in
event-by-event hydrodynamics based on the initial data, in much the way that
$\varepsilon_3$ predicts the orientation and magnitude of the leading
$v_3(p_T)$.  

As a first step, in \Sect{fourierspace} we correlate the principal momentum
space fluctuations with the Fourier modes of the geometry.  Based on this
analysis in \Sect{optimal} we construct a good geometrical predictor for the
orientation angle and magnitudes of the leading and subleading flows based on
two Fourier modes.

For comparison, we also correlate the subleading flow with a linear combination
of the complex $\varepsilon_{3,3}$ and $\varepsilon_{3,5}$,
\begin{subequations}
    \label{eps3all}
    \begin{align}
        \varepsilon_{3,3} \equiv   -\frac{[r^3 e^{i3 \phi }]}{ R_{\rm rms}^3 } \, , \label{eps33} \\
        \varepsilon_{3,5} \equiv   -\frac{ [r^5 e^{i3 \phi }]}{ R_{\rm rms}^5}  \, .\label{eps35} 
    \end{align}
\end{subequations} 
where the square brackets $[\,]$ denote an average over the initial entropy density
in a specific event, and $R_{\rm rms}=\sqrt{\llangle[r^2]\rrangle}$ is the
event averaged root-mean-square radius. Note that our definitions of
$\varepsilon_{3,3}$ and $\varepsilon_{3,5}$ are chosen to make 
the event-by-event quantities $\varepsilon_{3,3}$ and $\varepsilon_{3,5}$ 
linear in the fluctuations, since the denominator is a constant event-averaged quantity. In this respect this definition is different from the
conventional one which is a nonlinear function of the initial 
perturbations%
\footnote{We compared analogous results with $\varepsilon_{3,3}$ and
$\varepsilon_{3,5}$ defined via cumulants~\cite{Teaney:2010vd}, and found them
marginally worse than the ones presented in this paper.}
   (see \Sect{simulations} for further
explanation). 
We find that the subleading mode is also reasonably correlated with
a linear combination of these two quantities, but the quality of this predictor 
is considerably worse than a predictor based on two specific Fourier modes.

The geometric predictors described above are ultimately based on the
assumption of linear response. At least for the third harmonic (the scope of
this paper), these assumptions are checked in \Sect{linear_response}. In this
section we explicitly
compare the response to the average (``single-shot" hydrodynamics \cite{Qiu:2011iv}) and
the average  response (event-by-event hydrodynamics). We find reasonable
agreement between these two computational strategies for both the leading
and subleading triangular modes.


\section{PCA of Triangular Flow in Event-by-Event Hydrodynamics}
\label{PCA_sect}

\subsection{Principal components}
PCA was recently introduced in Ref.~\cite{Bhalerao:2014mua} (which includes one of
the authors) to quantify the dominant momentum space fluctuations of harmonic
flows in transverse momentum and rapidity in a precise way.  This section
provides a brief review of this statistical technique.

Paraphrasing Ref.~\cite{Bhalerao:2014mua}, in the flow picture of heavy ion
collisions the particles in each event are drawn independently from a single
particle distribution which  fluctuates from event to event.  The
event-by-event single particle distribution is expanded in a Fourier series
\st
\frac{\dd N}{\dd \p}   = V_0(p_T) + \sum_{n=1}^{\infty} V_n(p_T) e^{-in 
\varphi}  + {\rm H.c.}\, ,
\stp
where $\dd\p =  \dd y \,\dd p_{T} \, \dd\varphi$ notates the phase space,
$\varphi$  is the azimuthal angle of the distribution, and H.c. denotes
Hermitian conjugate. $V_n(p_T)$ is a complex Fourier coefficient recording the
magnitude and orientation of the $n$th harmonic flow.  This definition
deviates from the common practice of normalizing the complex Fourier
coefficient by the multiplicity, $v_n(p_T) = V_n(p_T)/V_0(p_T)$.

Up to non-flow corrections of order the multiplicity $N$, the long-range part
of the two-particle correlation function is determined by the statistics of the
event-by-event fluctuations  of the single distribution
\st
\llangle \frac{\dd N_\text{pairs}}{\dd\p_1 \dd \p_2 }  \rrangle  = \llangle \frac{\dd N}{\dd \p_1} \frac{\dd N}{\dd \p_2} \rrangle + \mathcal O\left(N\right) \, .
\stp
If the two-particle correlation function  is also expanded in a Fourier series
\st
\llangle \frac{\dd N_\text{pairs}}{\dd\p_1 \dd \p_2 }  \rrangle  = \sum_{n} V_{n\Delta}(p_{T1}, p_{T2}) e^{-in(\varphi_1 - \varphi_2)} \, , 
\stp
then this series determines the statistics of $V_n(p_T)$
\st
   V_{n\Delta}(p_{T1}, p_{T2}) = \llangle V_n(p_{T1}) V^*_n(p_{T2}) \rrangle \, .
\stp
The covariance matrix  $V_{n\Delta}(p_{T1},p_{T2})$, which is real,  symmetric,
and positive-semidefinite, can be decomposed into real eigenvectors,
\begin{align}
   V_{n\Delta}(p_{T1}, p_{T2}) =&  \sum_{a} \lambda^{a} \psi^{(a)}(p_{T1}) \psi^{(a)} (p_{T2})\label{dec1}, \\
   =& \sum_{a} V_n^{(a)}(p_{T1}) V^{(a)}_n(p_{T2})\label{dec2},
\end{align}
where $V_n^{(a)}(p_T)\equiv\sqrt{\lambda^{a} } \,\psi^{(a)}_n(p_T)$ and
$\int_0^\infty\!dp_T\,\psi^{(a)}\psi^{(b)}=\delta_{ab}$.  As discussed
above we have not normalized $V_3(p_T)$ by the multiplicity.  To make contact
with previous work, we define and present numerical results for 
\st
\| v^{(a)}_n\|^2  \equiv \frac{\int \big(V_n^{(a)}(p_T)\big)^2dp_T}{\int 
\left<dN/dp_T\right>^2dp_T}= \frac{\lambda_a}{\int\! 
\left<dN/dp_T\right>^2dp_T}\label{def_vn} \, ,
\stp
which scales with multiplicity and $\varepsilon_3$  in the same way as an
integrated $v_{3}\{2\}$ measurement.  Typically in event-by-event hydrodynamics
(as shown below) the eigenvalues are strongly ordered,  and two
eigenvectors describe the variance in the harmonic flow to 0.1\% accuracy.
Thus, PCA provides a remarkably economical description of the momentum
dependence of flow  fluctuations.  

The harmonic flow in each event can be decomposed into its principal
directions,
\st
V_3(p_T) =  \xi_1 V^{(1)}_3(p_T) + \xi_2 V^{(2)}_3(p_T) +\ldots\label{def_xi}  \, .
\stp
The real vectors $V_3^{(1)}(p_T), V_3^{(2)}(p_T), \ldots$ (which do not
fluctuate from event to event)  record the root-mean-square amplitude of the
leading and subleading flows.  The complex coefficients $\xi_1, \xi_2,\ldots$
indicate the orientation and event-by-event amplitude of their respective
flows.  The amplitudes of the different components are uncorrelated by
construction 
\st
\llangle \xi_a \xi_b^*\rrangle =  \delta_{ab} \, .
\stp
The original impetus for this work was a desire to understand which aspects of
the geometry are responsible for the orientation angle of the second principal
component.

\subsection{Simulations}
\label{simulations}
In this paper we  use boost-invariant event-by-event hydrodynamics to study
the principal components of $V_3(p_T)$ for LHC initial conditions.  The
implementation details of the hydrodynamics code will be reported elsewhere,
and here we note only the most important features.  Our simulations are
boost invariant and implement second order viscous
hydrodynamics~\cite{Baier:2007ix}, using a code base which has been developed
previously~\cite{Dusling:2007gi,Teaney:2012ke}.  For the initial conditions we
use the Phobos Glauber Monte Carlo \cite{Alver:2008aq}, and we distribute the
entropy density in the transverse plane according to a two-component model.
Specifically, for the $i$th  participant we assign a weight
\st
A_i \equiv \kappa \left[ \frac{ (1- \alpha) }{2}   + \frac{\alpha}{2} (n_{\rm coll})_i \right] \, ,
\stp
with $\alpha=0.11$,  $\kappa=35.1$ for $\eta/s=0.08$, and $\kappa = 32.8$
for $\eta/s=0.16$. $(n_{\rm coll})_i$ is the number of binary collisions
experienced by the $i$th participant; so the total number of binary
collisions is $ N_{\rm coll} = \half \sum_{i} (n_{\rm coll})_i$.  The entropy
density in the transverse plane at initial time $\tau_o$ and transverse
position $\x=(x,y)$ is taken  to be
\st
s(\tau_o,\x)  = \sum_{i\element {\parts}} s_{i}(\tau_o, \x - \x_i) \, ,
\stp
where $\x_i = (x,y)$ labels the transverse coordinates of the $i_{\rm th}$ participant, and
\st
s_{i}(\tau_o, \x) = A_i \, \frac{1}{\tau_o (2\pi \sigma^2) } e^{- \frac{x^2}{2\sigma^2} -\frac{y^2}{2\sigma^2} } \, , 
\stp
with $\sqrt{2}\sigma = 0.7\,{\rm fm}$.
The parameters $\kappa$ and $\alpha$ are marginally different from  Qiu's
thesis \cite{Qiu:2013wca}, and we have independently verified that this choice
of parameters reproduces the average multiplicity in the event.%
\footnote{ More precisely we have verified that for these parameters
    hydrodynamics with averaged initial conditions reproduces $\left. \dd
    N_{\rm ch}/\dd \eta\right|_{\eta=0}$ as a function of centrality after all
    resonance decays are included.  Assuming that the ratio of the charged
    particle yield to the direct pion yield is the same as in the averaged
    simulations, the current event-by-event simulations reproduces $\dd N_{\rm
ch}/\dd \eta$.}

The equation of state is motivated by lattice QCD
calculations~\cite{Laine:2006cp} and has been used previously by Romatschke
and  Luzum~\cite{Luzum:2008cw}.  In this paper we  compute ``direct" pions
(i.e. pions calculated directly from the freeze-out surface) and we do not
include resonance decays. We use a freeze-out temperature of $T_{\rm fo} =
140\,{\rm MeV}$.

Simulation results were generated for fourteen 5\% centrality classes with
impact parameter up to $b=\SI{12.4}{fm}$ and at two viscosities, $\eta/s=0.08$
and $\eta/s=0.16$. Unless specified, the results are for $\eta/s=0.08$.  We
generated 5000 events per centrality class.%
\footnote{
We thank Soumya Mohapatra for collaboration during the initial stages of this
project.}
We then performed PCA for the third harmonic $V_3(p_T)$ by discretizing
$V_3(p_T)$ results from hydrodynamics into 100 equally spaced bins between
$p_{T} = 0\ldots 5\,{\rm GeV}$, and finding the eigenvalues and eigenvectors of
the resulting Hermitian matrix.  Similar results for the other harmonics will
be discussed elsewhere.

Table~\ref{glaubertable} records the Glauber data which is used in this
analysis.  Event-by-event averages with the initial entropy density are notated
with square brackets, e.g.
\st
[r^2]  \equiv \frac{1}{\overline S_{\rm tot} } \int d^2\x \, \tau_o s(\tau_o,\x) r^2,   
\stp
where $\overline S_{\rm tot}$ is the average total entropy in a given
centrality class, $\left<\int d^2\x \, \tau_o s(\tau_o,\x)\right>$.  Averages
over events are notated with $\llangle\,\rrangle$, so that the root mean square
radius is
\st
  R_{\rm rms}\equiv \sqrt{\left<[r^2]\right>}\, .
\stp
As a technical note, here and below the radius is measured from the center of
entropy, so $[\x]=0$.  $\varepsilon_{3,3}$ and $\varepsilon_{3,5}$ are defined
in a somewhat unorthodox fashion in \Eq{eps3all}, with $\varepsilon_{3,3}^{\rm
rms} \equiv \sqrt{\llangle |\varepsilon_{3,3}|^2  \rrangle}$.
$\overline{r}_{\rm max}$ is the averaged maximum participant radius, ${\rm max}
\, |\x_i| $.
\begin{table}
\begin{center}
\begin{tabular}{|c |c |c | c | c | c |} \hline
   $\vphantom{\overline{\overline{H}}}$ Centrality& ($b_{\rm min}, b_{\rm 
   max}$) & $\overline N_{\rm part}$ & $R_{\rm rms}$ & $\overline{r}_{\rm max}$ 
   & $\varepsilon_{3,3}^{\rm rms}$ \\ \hline \hline
0--5  \%  & (0.0, \,  3.3) & 384 &  4.1 &  8.1 & 0.11\\
5--10 \%  & (3.3, \,  4.7) & 335 &  3.9 &  7.8 & 0.14\\
10--15\%  & (4.7, \,  5.7) & 290 &  3.7 &  7.5 & 0.17\\
15--20\%  & (5.7, \,  6.6) & 250 &  3.6 &  7.3 & 0.20\\
20--25\%  & (6.6, \,  7.4) & 215 &  3.4 &  7.0 & 0.22\\
25--30\%  & (7.4, \,  8.1) & 184 &  3.3 &  6.7 & 0.25\\
30--35\%  & (8.1, \,  8.8) & 156 &  3.2 &  6.4 & 0.28\\
35--40\%  & (8.8, \,  9.4) & 132 &  3.1 &  6.2 & 0.32\\
40--45\%  & (9.4, \,  9.9) & 110 &  3.0 &  5.9 & 0.35\\
45--50\%  & (9.9, \, 10.5) & 91  &  2.9 &  5.7 & 0.39\\
50--55\%  & (10.5,\, 11.0) & 74  &  2.7 &  5.4 & 0.44\\
55--60\%  & (11.0,\, 11.5) & 60  &  2.7 &  5.1 & 0.48\\
60--65\%  & (11.5,\, 11.9) & 47  &  2.6 &  4.8 & 0.52\\
65--70\%  & (11.9,\, 12.4) & 37  &  2.4 &  4.4 & 0.55\\\hline
\end{tabular}
\end{center}
\caption{
   Table of parameters from the Glauber model (all distances are measured in 
   fm).
   \label{glaubertable}}
\end{table}

\section{Subleading Triangular Flow}
\label{subleading}

As a first step, we list the (scaled) magnitudes of flows $\|v_{3}^{(a)}\|$ 
[\Eq{def_vn}] in central
collisions for the simulations described above:   
	\begin{center}
		\begin{tabular}{|c ||c |c | c | c |} \hline
			$a$ & 1 & 2 & 3 & 4\\\hline\hline
			$\|v_3^{(a)}\|$ & \num{1.5e-2} & \num{2.6e-3} & \num{4.8e-4} & \num{1.1e-4} \\ \hline
		\end{tabular}
	\end{center}
   Note that the quantities in this table are proportional to the square-root of
   the eigenvalues, $\|v_3^{(a)}\| \propto \sqrt{\lambda_a}$.
From the decreasing magnitudes  of the listed (scaled) magnitudes, we see that
the first two eigenmodes account for 99.9\% of the squared variance, which can
be represented as a  sum of the eigenvalues
\begin{align}
   \int_0^{\infty} \dd p_{T} \llangle V_{3}(p_T)  V_{3}^*(p_T) \rrangle =&  
   \sum_{a}\lambda_a  \propto \sum_{a} \|v_3^{(a)}\|^2 \, .
\end{align}

\begin{figure*}
	\includegraphics{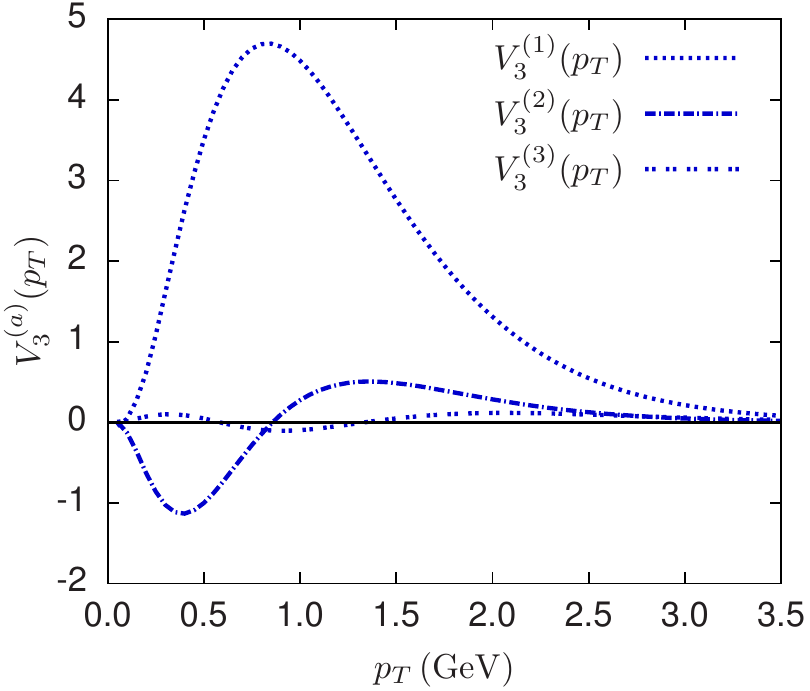}\qquad
	\includegraphics{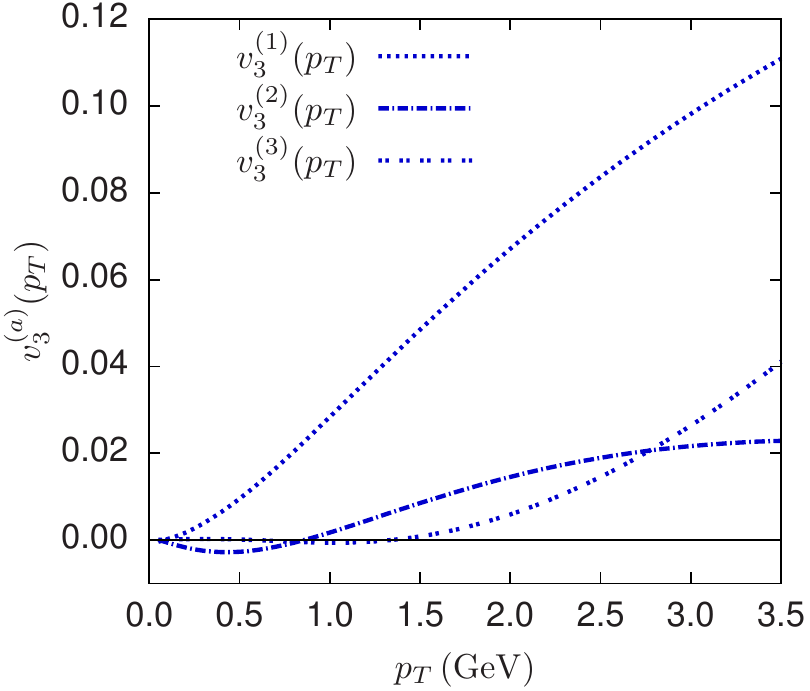}%
	\caption{Momentum dependence of flow components in central collisions. a) Principal flow vectors, $V^{(a)}_3(p_T)$. b) Principal flow vectors divided by 
		the average multiplicity, 
    $v_3^{(a)}(p_T) \equiv V_3^{(a)}(p_T) /\llangle \dd N/\dd p_T 
    \rrangle$.\label{PCA_Vn_1_nonorm}}
\end{figure*}

Figure \ref{PCA_Vn_1_nonorm}(a) displays the eigenvectors, $V_{3}^{(a)}(p_T)$, 
for the
leading and first two subleading modes.  We see that  only the first two flow
modes are significant, and in the rest of this paper we consider only
these two. To make contact with the more traditional definitions of $v_3(p_T)$,
we divide by $\llangle \dd N/\dd p_T \rrangle$ and present the same eigenmodes
in \Fig{PCA_Vn_1_nonorm}(b).

\begin{figure}
	\includegraphics{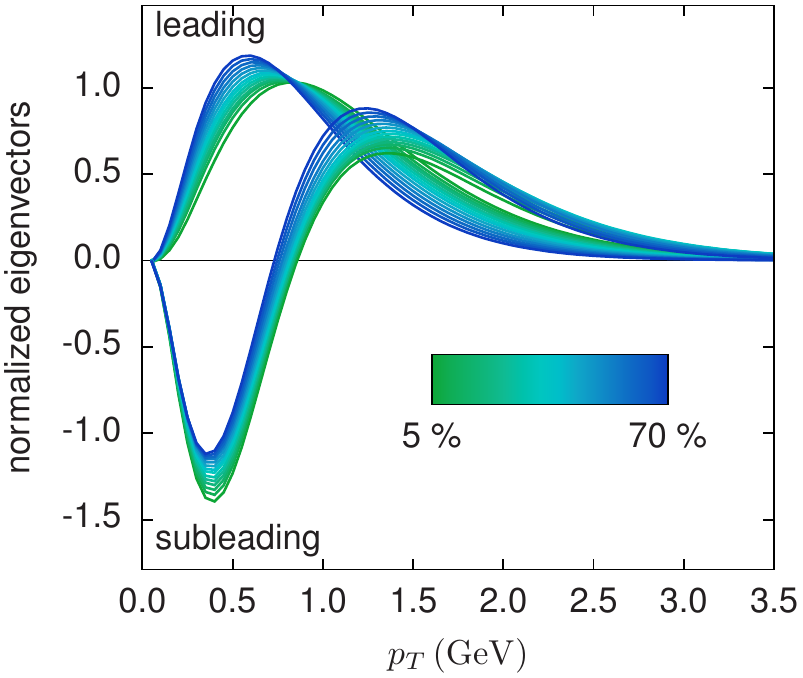}
	\caption{Centrality dependence of flow eigenvectors $\psi^a(p_T)$.\label{PCA_Vn_2}}
\end{figure}

We also investigated the  centrality and viscosity dependence of the principal
components. The normalized principal flow eigenvectors $\psi^a(p_T)$ are
approximately independent of viscosity (not shown). In \Fig{PCA_Vn_2}, we show
the centrality dependence of these normalized eigenvectors. In more central
collisions the eigenvectors shift to larger transverse momentum, which can be
understood with the system size scaling introduced in Ref.~\cite{Basar:2013hea}.

\begin{figure}
	\includegraphics{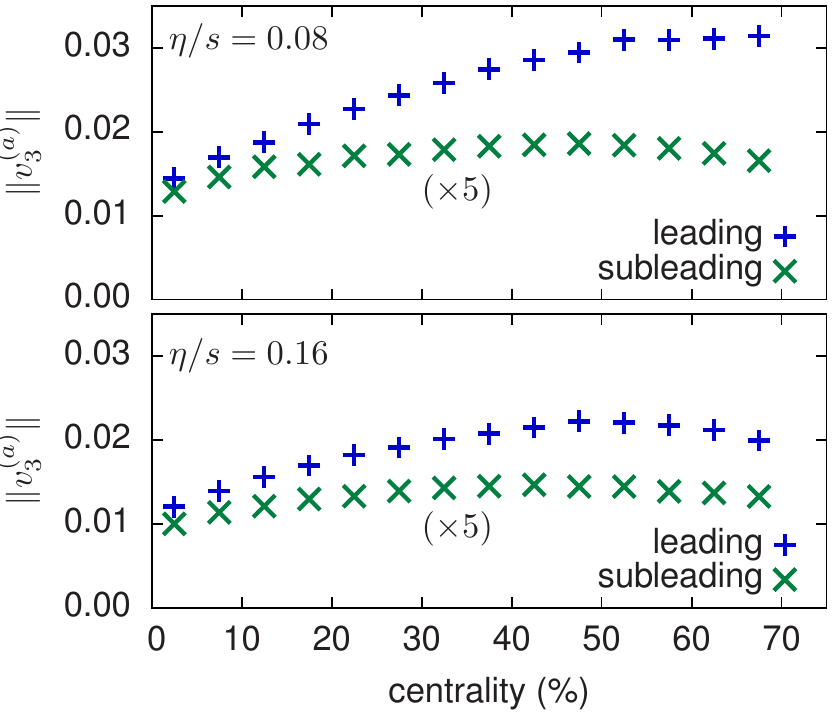}
	\caption{Centrality and viscosity dependence of scaled eigenvalues $\|v^{(a)}_3\|$. 
    (The subleading flow has been magnified 5 times to bring to scale with leading flow.)
\label{PCA_eval_2}}
\end{figure}

The magnitude of the flow, i.e. the squared integral $\int\,
(V_{3}^{(a)}(p_T))^2dp_T$, depends on both centrality and viscosity. To
factor out the trivial multiplicity dependence of $V_3(p_T)$,  we plot the
scaled flow eigenvalues $\|v_n^{(a)}\|$ [see \Eq{def_vn}] in \Fig{PCA_eval_2}.
Going from $\eta/s=0.08$ to $\eta/s=0.16$ we see significant suppression of the
leading mode.  In general the subleading scaled flow $\|v_3^{(a)}\|$ depends
weakly on centrality.

\section{Geometric Predictors for Subleading Flow}
\label{Geometry}
\subsection{Average geometry in the subleading plane}
\label{avg_geometry}

In this section, we clarify the physical origin of the subleading flow by
correlating the subleading hydrodynamic response with the geometry.

\begin{figure*}
\includegraphics{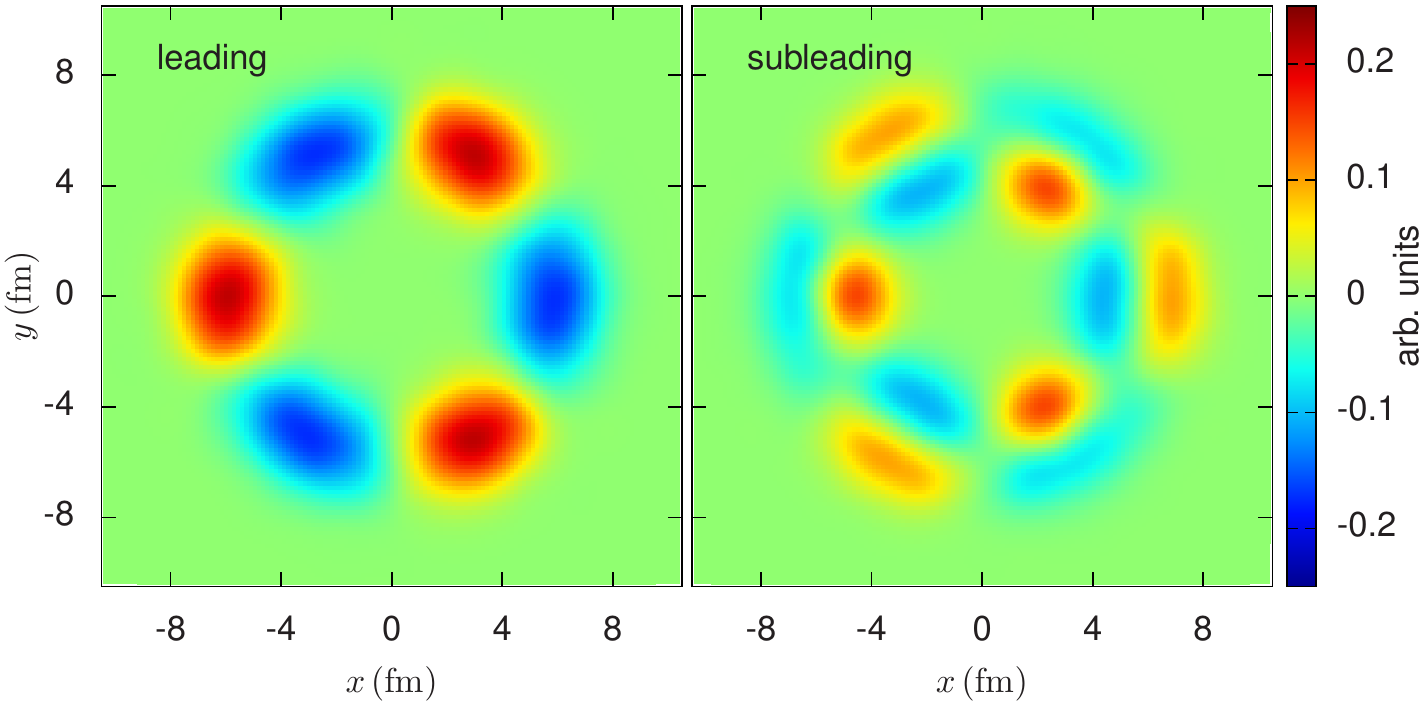}
\caption{Average geometry$\times r^3$ in the leading and subleading
 principal component planes in central collisions 
minus an averaged radially symmetric background, $r^3 (\overline{S}(\x;\,\xi_a) - \left<S(\x)|\xi_a|\right>)$.
Peak fluctuations are $\pm$10--20\% above the background.\label{PCA_S_1}}
\end{figure*}

As a first step, we determined the average initial geometry in the principal
component plane.  Specifically, for each event the phase of the principal
component $\xi_a$ [see \Eq{def_xi}] defines orientation of the flow. We then 
rotate each event
into $\xi_a$ plane and average the initial entropy density, $S(\x)\equiv \tau_o
s(\tau_o, \x)$. More precisely, the event-by-event geometry in the principal
component plane  is defined to be 
\begin{equation}
	S(r,\phi;\,\xi_a) \equiv \frac{1}{3}\sum_{\ell=0}^{2}S\left(r,\phi+(\arg \xi_a +2\pi \ell)/3\right) \, , 
\end{equation}
where we have averaged over the phases of $\sqrt[3]{\xi_a}$.  Next, we average
$S(r,\phi;\,\xi_a)$ over all events weighted by the magnitude of the flow
\begin{equation}
	\overline{S}(r,\phi;\,\xi_a) \equiv\left<S(r,\phi;\,\xi_a) |\xi_a|\right>\label{Srphibar}.
\end{equation} 
Figure \ref{PCA_S_1} shows the in-plane averaged geometry
$\overline{S}(r,\phi;\,\xi_a)$ for the leading and subleading principal
components in central collisions.  
Clearly, the leading principal component $V_{3}^{(1)}$ is strongly correlated
with the triangular components of the initial geometry, while the subleading
component $V_{3}^{(2)}$ is correlated with the radial excitations of this
geometry.

\begin{figure}
\includegraphics{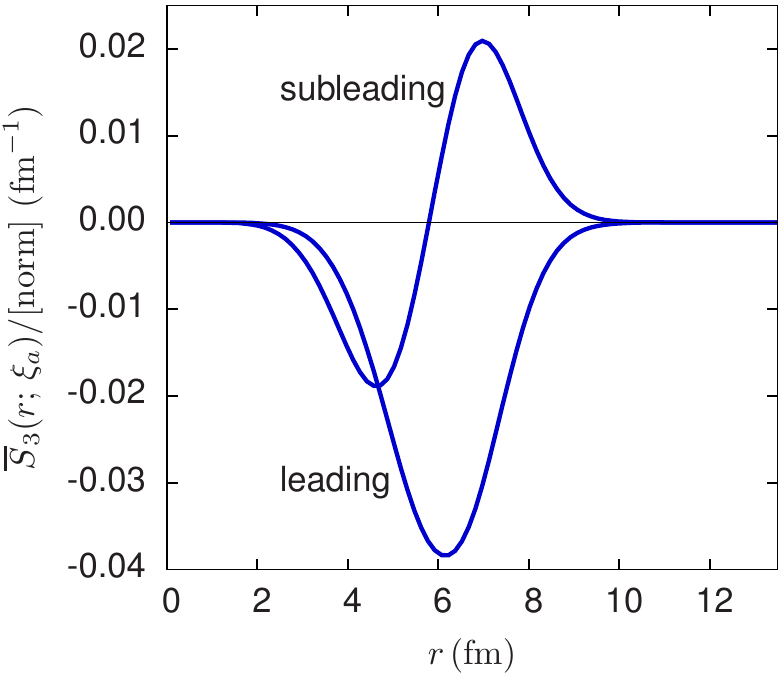}
\caption{\label{PCA_S_2} 
Correlation between the principal components and the 
triangular geometry, $\llangle S_3(r) \xi_a^* \rrangle$, for
the leading and subleading flows in central collisions.
The result has been multiplied by $r^4$ and normalized by $\overline{S}_{\rm tot}R_{\rm rms}^3$, so that the area under the leading curve is approximately $\varepsilon_{3,3}^{\rm rms}$.}
\end{figure}

To give a one-dimensional projection of \Fig{PCA_S_1}, we integrate
\Eq{Srphibar} over the azimuthal angle to define
\begin{equation}
	\overline{S}_3(r;\,\xi_a)\equiv \int_0^{2\pi}\!\!\!\text{d}\phi\, \overline{S}(r,\phi;\,\xi_a)\,e^{i3\phi}\label{Srbar}.
\end{equation}
This is equivalent to defining $S_3(r)$,
\st
S_{3}(r) \equiv \int_0^{2\pi} \!\!\! \dd \phi \, S(r,\phi) \,  e^{i3\phi}  \, , 
\stp
and correlating this with the flow fluctuation $\xi_a$
\st
\overline{S}_3(r;\,\xi_a) = \llangle S_3(r) \xi_a^* \rrangle \, .
\stp
Results for $\overline{S}_3(r;\,\xi_a)r^4$ are shown by the blue (gray) curves 
in
\Fig{PCA_S_2}. 

Again we see that the leading flow originates from a geometric fluctuation with
a large integrated eccentricity, while the subleading flow is sensitive to the
radial excitation of the triangularity. Note that the  relatively small
subleading flow corresponds to a fairly significant fluctuation of the initial
geometry.

\subsection{The average geometry in Fourier space}
\label{fourierspace}

It is evident from  \Fig{PCA_S_2} that the leading and subleading geometries
have different characteristic wave numbers. This becomes apparent when we
correlate the flow signal with the Fourier (or Hankel) transform of the
triangular geometry%
\begin{align}
   S_3(k)
   \equiv &\int_0^{\infty} \!r\dd r\, S_3(r)J_3(k r) \, .
\end{align}
Here $S_m(k)$ has the meaning of the $m$th harmonic of the 2D Fourier 
transform of the initial geometry $S(\k)$, i.e. 
$S_m(k)\equiv\frac{i^m}{2\pi}\int\!\dd\phi\, e^{im\phi_k}S(\k)$.

We recall that the $\varepsilon_{3,3}$ is determined by the long wavelength
limit of $S_3(k)$ \cite{Teaney:2010vd} 
\st
\lim_{k \rightarrow 0 } S_{3}(k) = -\overline{S}_{\rm tot} \frac{(k R_{\rm rms}/2)^3}{3!}  \varepsilon_{3,3} \, ,
\stp
where  $R_{\rm rms}$ is the root mean square radius and $\overline{S}_{\rm
tot}$ is the total entropy in a given centrality bin. The constant factors are
determined by the expansion of $J_3(x)$ near $x=0$.  Motivated by this limit we
define a generalized eccentricity $\varepsilon_3(k)$
\st
\varepsilon_3(k) \equiv 
-\frac{1}{\overline{S}_{\rm tot}} \int_0^{\infty} r dr S_3(r) \left[ \frac{3!}{(k{R}_{\rm rms}/2)^3}  J_3(kr) \right]\label{ek},
\stp
which approaches $\varepsilon_{3,3}$ as $k \rightarrow 0$.  Clearly in a
Glauber model there is an analogous definition
\st
\varepsilon_3(k) = -\frac{1}{\overline{N}_{\rm part}} \sum_{i=1}^{N_{\rm part}} e^{i3\phi_i} \left[ \frac{3!}{(k {R}_{\rm rms}/2)^3} J_3(k r_i)  \right] \, ,
\stp
where the coordinates of the $i$th participant are $\x_i = (r\cos\phi_i, 
r\sin\phi_i)$.

The Pearson correlation coefficient between the flow and a specific wave number
$k$ is 
\begin{equation}
   Q_a(k) \equiv \frac{ \llangle \xi_a \varepsilon_3^*(k)\rrangle }{\sqrt{ \llangle |\varepsilon_3(k) |^2 \rrangle \llangle |\xi_a|^2 \rrangle } } .
\end{equation}

\begin{figure}
\includegraphics{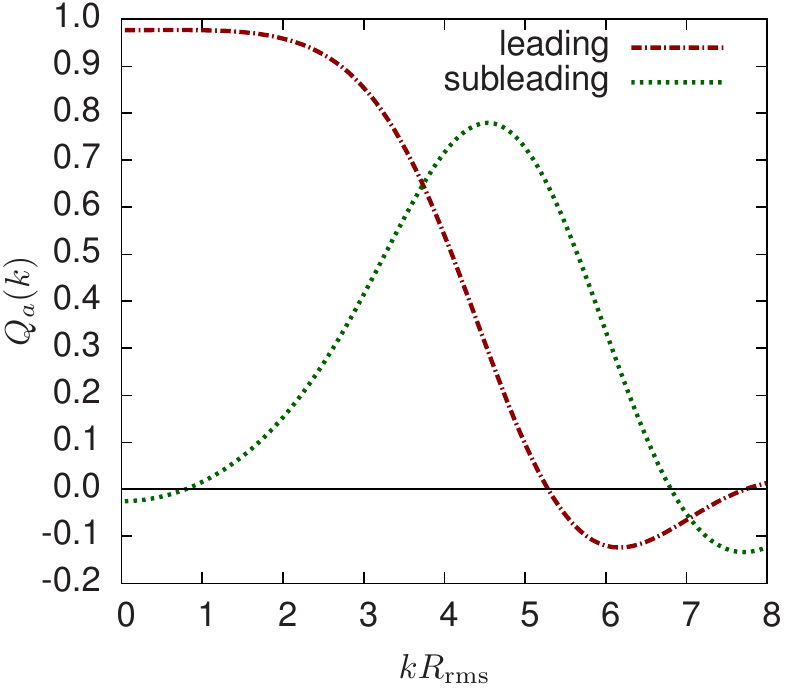}
\caption{Quality plot (or Pearson correlation coefficient)  for $\varepsilon_3(k)$ 
as a single-term predictor for principal flows (central collisions). 
\label{oneterm}}
\end{figure}
Examining $Q_a(k)$ in  \Fig{oneterm}, we see that leading component is produced
by low-$k$ fluctuations, while subleading flow originates from fluctuations at
larger $k$.

\subsection{Optimal geometric predictors for the subleading flow}
\label{optimal}

In this section, our aim is to predict the magnitude and orientation of the
leading and subleading flows.  To this end we  regress each principal
component of the flow with various Fourier components of the initial geometry.

Following Ref.~\cite{Gardim:2011xv}, we construct a prediction for the flow amplitude
$\xi_a^\text{pred}$ by taking a linear combination of $\varepsilon_3(k)$ :
\begin{equation}
   \xi^\text{pred} = \sum_{i=1}^{n_k} \omega_b  \varepsilon_3(k_b) \label{predictor}\, .
\end{equation}
The selected wave numbers $k_b$ are discussed in the next paragraph.  The
response coefficients  $\omega_b$ are chosen to minimize the square error
${\mathcal E}^2_a$, or equivalently to maximize the Pearson correlation
coefficient $Q_a$ between the flow and the prediction 
\begin{align}
   &\text{min}\quad	{\mathcal E}^2_a=\big<|\xi_a-\xi^\text{pred}_a|^2\big> \, , \\
&\text{max}\quad	Q_a=\frac{\big<\xi_a {\xi^*_a}^{\text{pred}}\big>}{\sqrt{\big<\xi_a \xi^*_a\big>\big<{\xi_a}^\text{pred} {\xi^*_a}^{\text{pred}}\big>}}\label{pears} \, .
\end{align}
The correlation coefficient $Q_a$ is referred to as the quality coefficient in
Ref.~\cite{Gardim:2011xv}.  

We  construct two predictors based on two and five wave numbers.  For the
five-term predictor we choose  equidistant points which span the range seen in
\Fig{oneterm} \st
k_b R_{\rm rms} = 1,\, 3,\, 5,\, 7, \,9 \, ,
\stp
and fit the response coefficients $\omega_b$.  The two term predictor was
motivated by the discrete Fourier-Bessel series advocated for
in Ref.~\cite{Floerchinger:2014fta},
\st
k_{b} R_{o} = j_{3,1}, \;  j_{3,2}   \qquad R_o \simeq 3\, R_{\rm rms}\, ,\label{twok}
\stp
where $(j_{3,1}, j_{3,2}) \simeq (6.38,9.76)$ are the zeros of $J_3(x)$, and we
select $R_o$ to optimize the correlation between the geometrical predictor and
the flow.  For comparison, we also constructed a two term linear predictor from
the familiar eccentricities $\varepsilon_{3,3}$ and $\varepsilon_{3,5}$ defined
in Eqs.~(\ref{eps33}) and (\ref{eps35}).

\begin{figure*}
	\includegraphics{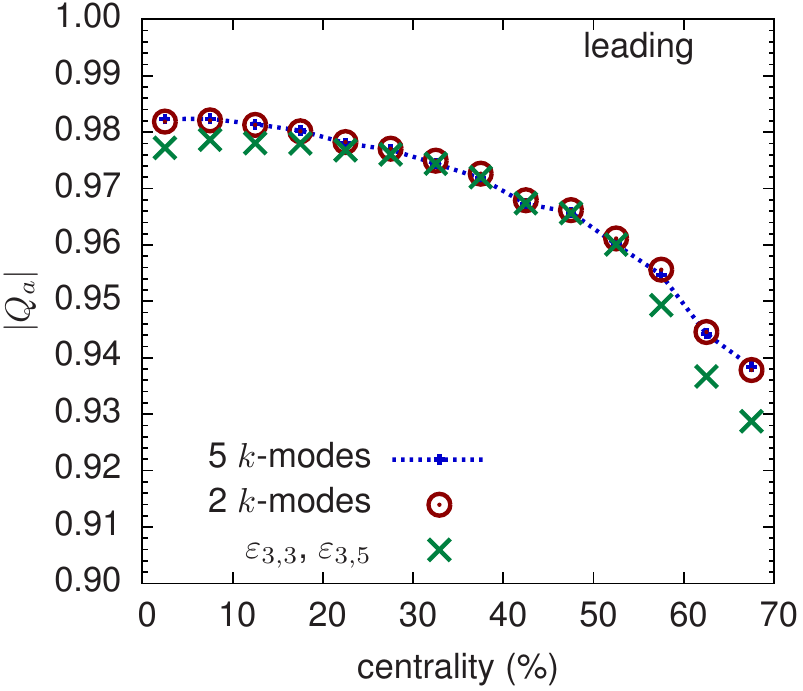}
	\includegraphics{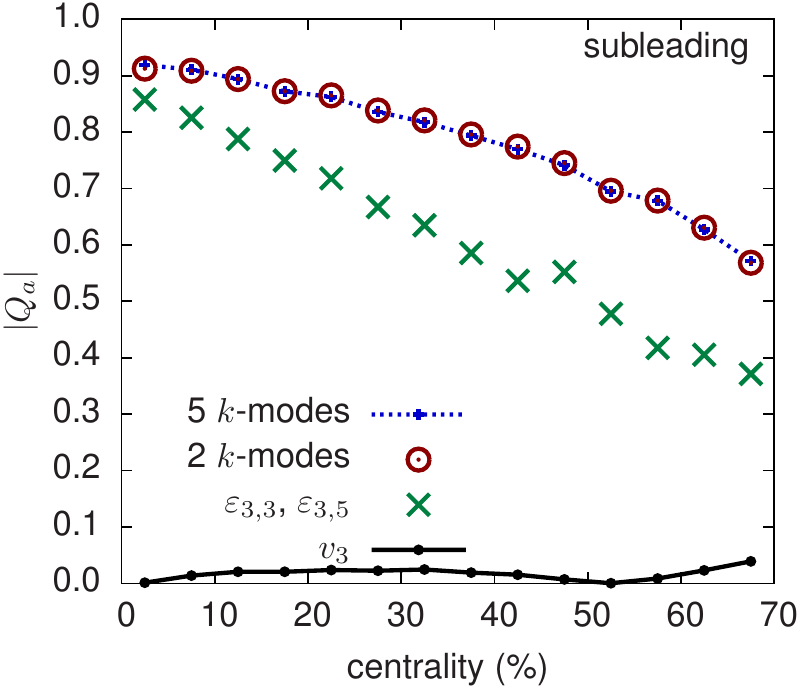}
    \caption{Pearson correlation coefficient for 2 and 5 $k$-mode predictors
    [\Eq{ek}], and a predictor based on the  eccentricities $\varepsilon_{3,3}$
and $\varepsilon_{3,5}$.  (a) Correlations for the leading flow (zero suppressed
for clarity). (b) Correlations for the subleading flow. \label{corrpsi}}
\end{figure*}

In Figs.~\ref{corrpsi}(a) and (b) we show the correlation coefficient between 
the
leading and subleading flow amplitudes $\xi_a$ and the predicted amplitudes
$\xi_a^{\rm pred}$ using the two and five Fourier mode fits, and the
$\varepsilon_{3,3}$, $\varepsilon_{3,5}$ fit.  As is well known, the leading
mode is very well predicted  by $\varepsilon_{3,3}$ and $\varepsilon_{3,5}$,
though the quality degrades towards peripheral collisions.  These results for
the leading mode can be  compared profitably with Fig. 4 of
Ref.~\cite{Gardim:2014tya}, where similar results were recently reported.  For the
subleading flow the linear correlation coefficient is reduced relative to the
leading flow, and a high degree of correlation is only achieved for the 0--40\%
centrality range.  The simple geometric predictor based on $\varepsilon_{3,3}$
and $\varepsilon_{3,5}$  is reasonably correlated with the subleading flow in
central collisions,  but this correlation rapidly deteriorates in more
peripheral collisions.  The predictor based on two judiciously chosen wave
numbers generally outperforms all other predictors we studied for both the
leading and subleading modes.  Indeed, we believe based on numerous other fits
that these two wave numbers essentially exhaust the predictive power of the
event-by-event triangular geometry, $S_3(r)$.   Additional histograms
correlating the amplitudes and phases of the flow and the two-term prediction
are shown in \Fig{phi} in the appendix.

The two-wave-number  fit correlates the flow with a specific projection of the
triangular geometry, i.e.
\st
\xi_a^{\rm pred} \propto \int_0^{\infty} r dr S_{3}(r)  \rho(r) \label{rho}\,  ,
\stp
where $\rho(r)$ is a radial weight chosen to maximize the correlation between
the flow and the projection.  This is analogous to  using $\varepsilon_{3,3}$
to predict triangular flow, where the radial weight is $\rho(r)\propto-r^3$
\st
\varepsilon_{3,3} \propto -\int_0^{\infty} r dr S_3(r) r^3 \, .
\stp
We have used Fourier modes as a basis for $\rho(r)$, 
\st
\rho(r)  \propto \omega_1 \frac{J_3(k_1 r)}{(k_1 R_{\rm rms})^3} + \omega_2 \frac{J_{3}(k_2 r)}{(k_2 R_{\rm rms})^3}\label{basis} \, ,
\stp
but other functions could have  been used.\footnote{A table of
$\omega_1/\omega_2$ is given as a function of centrality in
the appendix.}  In \Fig{projector} we compare the radial weights for
the leading and subleading modes.  The overall normalization of weight function
$\rho(r)$ is adjusted so that
\begin{equation}
\left<\left|\int_0^\infty\!r\\d r \rho(r)S_3(r)\right|^2\right>=S_{\rm tot}^2.\label{rhonorm}
\end{equation}
The weight function for the leading projector is very close to cubic weight,
but the subleading radial weight has a node at $r\simeq 1.5\,{R}_{\rm rms}$ .
\begin{figure}
	\includegraphics{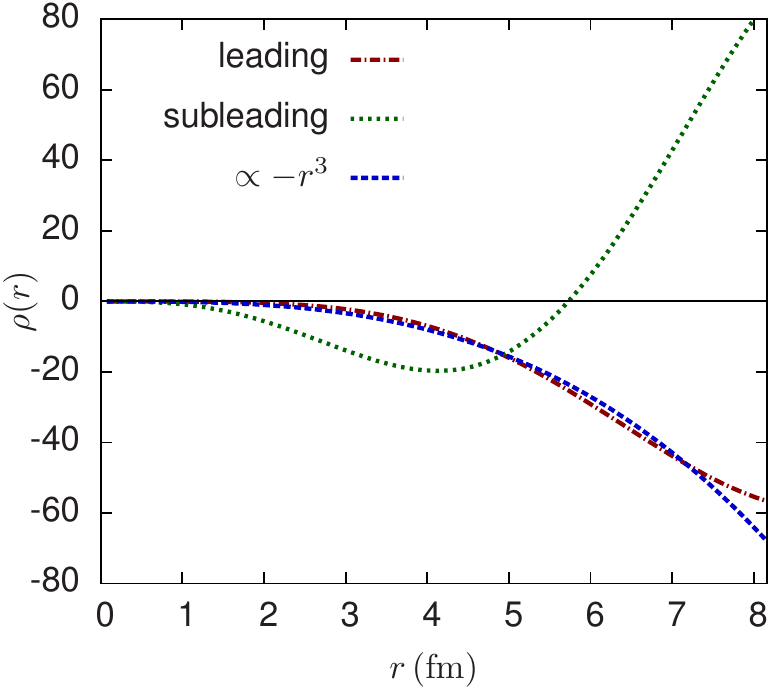}
	\caption{Radial weight functions for the leading and subleading flow predictors in central collisions, see \Eq{rho}.\label{projector}}
\end{figure}

\begin{figure}
	\includegraphics{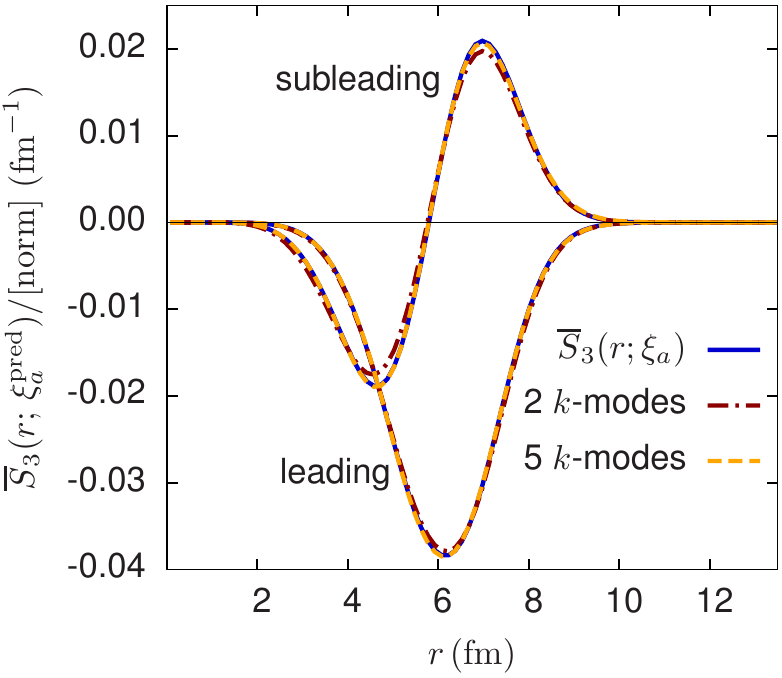}
	\caption{Comparison of the averaged geometry in the principal component 
	plane, $\llangle S_3(r) \xi_a^* \rrangle$, and two- and five-term predictor 
	plane,  $\llangle S_3(r) \xi_a^{*\text{pred}} \rrangle$, in central 
	collisions.
		\label{PCA_En_1}
	}
\end{figure}

Within the framework of linear response, in \Sect{avg_geometry} we found the
optimal geometry for predicting  the leading and subleading flows by
correlating the observed flow with the geometry, $\llangle S_3(r)
\xi_a^*\rrangle$.  To test if the two and five wave number predictors reproduce
this optimal geometry, we formed the analogous correlator between the predicted
flow $\xi^{\rm pred}$ and $S_3(r)$, $\llangle S_3(r) \xi^{*\rm pred}_a
\rrangle$.  Examining \Fig{PCA_En_1}, we see that the two term predictor fully
captures the optimal average geometry.  For peripheral collisions the optimal
geometry differs from what we can construct using linear combinations of
Fourier modes, suggesting that additional nonlinear
physics~\cite{Qiu:2011iv,Gardim:2011xv,Teaney:2012ke} plays a role in
determining the subleading flow.

Figure \ref{corrpsi}(b) also shows the correlation (or lack thereof)
between the subleading flow and the integrated $v_3$
\st
  Q \equiv \frac{\left<\xi_2 v_3^*\right>}{\sqrt{\left<v_3v_3^*\right>\left<\xi_2\xi_2^*\right>}} \, .
\stp
Since the subleading mode is uncorrelated with the leading mode (by
construction), there is almost no correlation between the integrated $v_3$ and
the subleading mode.  The upshot is that measurements of $v_3(p_T)$ based on
the event plane or scalar product method are projecting out the important
physics of the subleading mode.

\section{Testing Linear Response}
\label{linear_response}

The success of the linear flow predictors discussed in previous section depends
on the applicability of linear response.  A straightforward way to check this
assumption is to compare the averaged response of event-by-event hydrodynamics
to the  hydrodynamic response to suitably averaged initial conditions.

\begin{figure}
\includegraphics{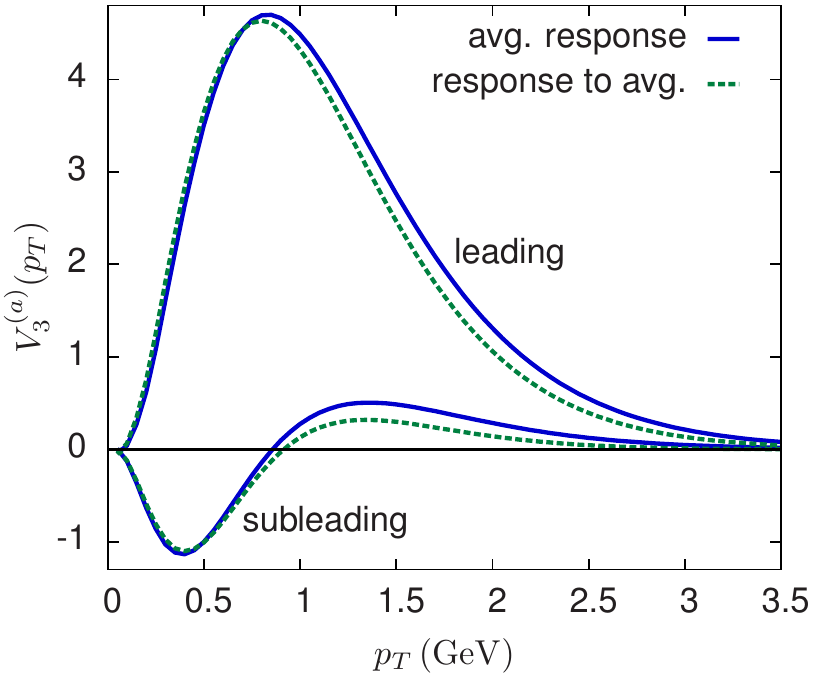}
\caption{Comparison of event-by-event hydro (averaged response) and single-shot hydrodynamics
(response to average geometry) in central collisions. The singe-shot hydrodynamic results are generated
from the initial conditions in \Fig{PCA_S_1}\label{PCA_Vn_res_1}.}
\end{figure}

In \Sect{avg_geometry} we computed the  average geometry in the event planes of
the leading and subleading flows (see \Fig{PCA_S_1}).  It is straightforward to
simulate this smooth initial condition and to compute the associated
$V_{3}(p_T)$. This is known as ``single-shot" hydrodynamics in the
literature~\cite{Qiu:2011iv}.  In \Fig{PCA_Vn_res_1}  we compare  $V_{3}(p_T)$
from the leading and subleading average geometries to the principal components
$V_{3}^{(1)}(p_T)$ and $V_{3}^{(2)}(p_T)$ of event-by-event hydro.
The qualitative features of both principal components are reproduced well by
single-shot hydrodynamics, especially for the leading flow.  It is particularly
notable how the single-shot evolution reproduces the  change of sign in
$V_3^{(2)}(p_T)$.  However,  in an important $p_T$ range, $p_T\sim 1.2 \,{\rm
GeV}$, the single-shot evolution misses the event-by-event curve for subleading
flow by $\sim30\%$.

\begin{figure*}
\includegraphics[scale=1.0]{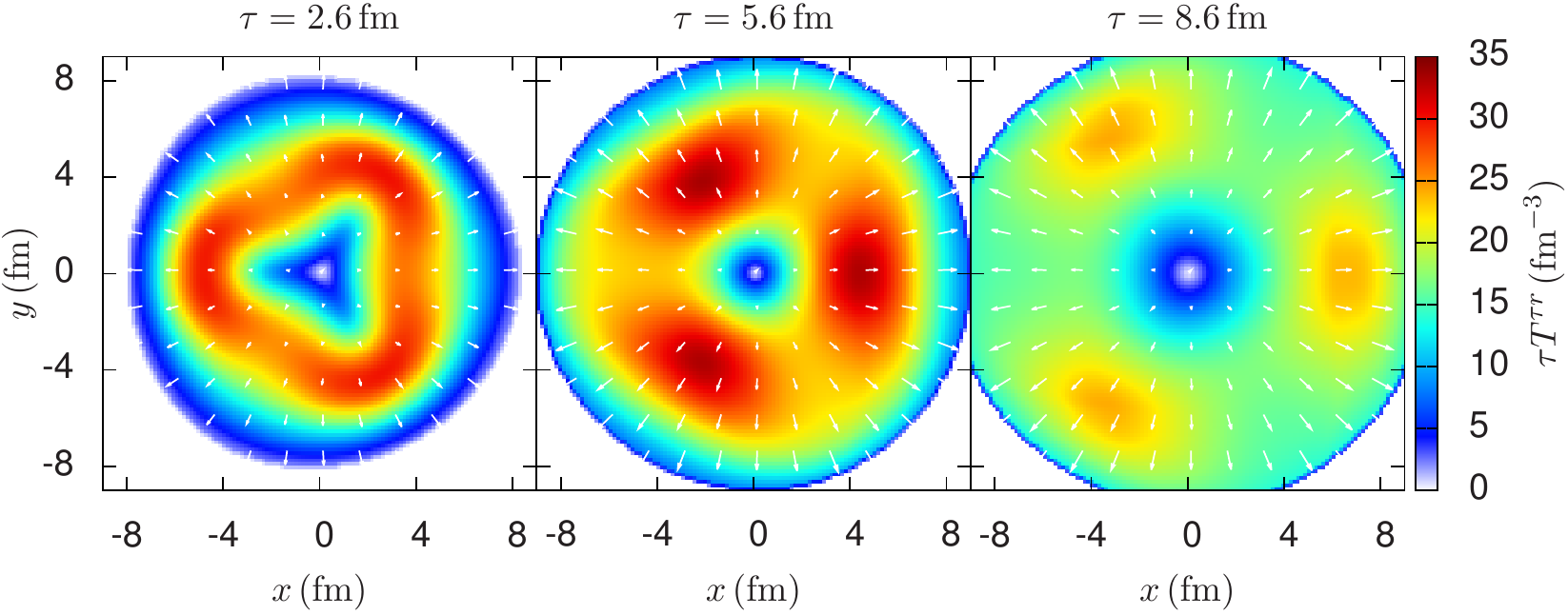}
\caption{ Hydrodynamic evolution of the subleading triangular flow for the
averaged initial conditions shown in \Fig{PCA_S_1}(b).  The color contours
indicate the radial momentum density per rapidity, $\tau T^{\tau r}$, while the
arrows indicate the radial flow velocity.\label{evolv}}
\end{figure*}

It is useful to examine the time development of the subleading flow in the
single-shot hydrodynamics.  In \Fig{evolv}, we present three snapshots of the
subleading flow evolution. The color contours show the radial momentum density
per rapidity,
\begin{equation}
	\tau T^{\tau r}=\tau (e+p) u^\tau u^r \,, \label{tauT0r}
\end{equation} 
as a function of proper time $\tau$.

Shortly after the formation of the fireball, at $\tau = \SI{2.6}{fm}$ we
observe negative triangular flow in \Fig{evolv}(a). This flow is produced by
the excess of material at small radii flowing into the ``valleys" at larger
radii [see \Fig{PCA_S_1}(b)]. However, the radial flow has not developed yet,
and therefore this phase of  the evolution creates negative flow at small
transverse momentum.  After this stage, we see typical flow evolution of a
triangular perturbation, i.e. the negative geometric eccentricity at small
radii is transformed into positive triangular flow at large transverse momentum
[see Figs.~\ref{evolv}(b) and (c)]. The inner eccentricity dominates over the 
outer
eccentricity at high $p_T$ because the radial flow has more time to develop
before freeze-out, and because there is more material at small radii.


\section{Discussion}
\label{discussion}

This paper illustrates how principal component analysis can be used to
understand the physics encoded in the two particle correlation matrix of
hydrodynamics.  PCA is an  economical way to summarize the factorization
breaking in these correlations.  More precisely, we found that  the $r$ matrix
of correlation coefficients in hydrodynamics, \Eq{rmatrix}, is completely
described by two principal components, $V_3^{(1)}(p_T)$ and $V_3^{(2)}(p_T)$ as
written in \Eq{pcar}.  Importantly, these components have a simple physical
interpretation---they  are the hydrodynamic response to two statistically
independent initial conditions in the fluctuating geometry.  The leading
principle component is the hydrodynamic response to the participant
triangularity, while the subleading flow (which is uncorrelated with the
leading flow) is the hydrodynamic response to the first radial excitation of
the triangularity.  This conclusion was reached by averaging the event-by-event
geometry in the event plane of the subleading flow (\Fig{PCA_S_1}).  The
magnitude of this radial excitation is on par with the  magnitude of the
triangularity (\Fig{PCA_S_2}), although the hydro response is smaller in
magnitude.  Since the subleading component is uncorrelated with the integrated
$v_3$, it is projected out in analyses of triangular flow based on the scalar
product or event plane methods.

We first studied the basic properties of the subleading flow such as its
dependence on transverse momentum (\Fig{PCA_Vn_1_nonorm}), and centrality and
shear viscosity (\Fig{PCA_eval_2}).  The flow response is approximately linear
to the  geometrical deformation.  This was checked by simulating the response
to the average in-plane geometry with ``single-shot" hydrodynamics
(\Fig{evolv}), and comparing this result to event-by-event hydrodynamics;
i.e., we compared the response to the average with the averaged response
(\Fig{PCA_Vn_res_1}).

Motivated by the linearity of the response, we constructed a geometrical
predictor for the subleading flow analogous to $\varepsilon_{3,3}$.  We first
defined $\varepsilon_3(k)$ as the $k{-}{\rm th}$ Fourier mode of the
event-by-event triangular geometry up to normalization.\footnote{ The
normalization is chosen so that $\lim_{k\rightarrow 0} \varepsilon_3(k)  =
\varepsilon_{3,3}$.  }  Then we constructed a linear geometrical  predictor
for the leading and subleading flow angles and magnitudes based on two Fourier
modes (Figs.~\ref{corrpsi} and \ref{phi}).  Indeed, the subleading flow 
response is
proportional to an event-by-event quantity which captures the radial excitation
of the triangular geometry, \st \int  \dd^2\x   \; s(\tau_o, \x) \, e^{i3\phi}
\rho(r) \, , \stp where $s(\tau_o,\x)$ is the initial entropy distribution,
and $\rho(r)$ is an appropriate excited radial weight function.  The two term
Fourier fit to $\rho(r)$ is tabulated in the appendix and graphed in
\Fig{projector}.  The subleading flow probes the initial state geometry at
higher wave numbers than the leading flow (\Fig{oneterm}). We found that the
correlation between the flow and the Fourier components of the geometry is
maximized for wave numbers away from zero, $k R_{\rm rms} \sim 4.0$.  Thus, the
subleading flow provides a new test of  viscous hydrodynamics and initial-state
models.

In peripheral collisions the correlation between the linear geometrical
predictor and the  flow is smaller. This suggests that nonlinear dynamics at
large impact parameters couples the average elliptic geometry to the harmonic
perturbations~\cite{Qiu:2011iv,Gardim:2011xv,Teaney:2012ke}.  The statistical
tools such as PCA and related methods developed in this work can be used to
clarify this complex hydrodynamic response. \\

\noindent{ \bf Acknowledgments:} \\
{}\\
We thank J.~Y.~Ollitrault, E.~Shuryak, and J.~Jia for continued interest.  We
especially thank S.~Mohapatra for simulating hydro events.  This work was supported by the Department of Energy,
DE-FG-02-08ER41450. 

\appendix


\section{Two term predictor}
\label{twotermpredictor}
Here we present the best fit results for the two wave number predictor, see
eqs.~(\ref{rho}), (\ref{basis}) and (\ref{rhonorm}),
\st
k_{b} R_{o} = j_{3,1}, \;  j_{3,2}   \qquad R_o \simeq 3\, R_{\rm rms}\, .
\stp
Table~\ref{twotermtable} records the ratios of fit coefficients
$\omega_2/\omega_1$ for the leading and subleading predictors.
\begin{table}
	\begin{center}
		\begin{tabular}{| c | c | c|}\hline
			centrality & \parbox[c]{1.5cm}{leading \\ $\omega_2/\omega_1$} & \parbox[c]{1.5cm}{subleading $\omega_2/\omega_1$} \\ \hline\hline
0-5\%    &-0.93 & -2.61\\
5-10\%   &-0.87 & -2.89\\
10-15\%  &-0.83 & -3.03\\
15-20\%  &-0.78 & -3.08\\
20-25\%  &-0.73 & -3.16\\
25-30\%  &-0.70 & -3.18\\
30-35\%  &-0.63 & -3.22\\
35-40\%  &-0.58 & -3.17\\
40-45\%  &-0.53 & -3.15\\
45-50\%  &-0.44 & -3.12\\
50-55\%  &-0.35 & -3.07\\
55-60\%  &-0.23 & -3.02\\
60-65\%  &-0.06 & -2.99\\
65-70\%  & 0.08 & -2.88\\
            \hline
		\end{tabular}
	\end{center}
	\caption{
		Table of two term predictor coefficients,  eqs.~(\ref{rho}), (\ref{basis}) and (\ref{rhonorm}).
		\label{twotermtable}}
\end{table}

\begin{figure*}
   \includegraphics[width=0.6\textwidth]{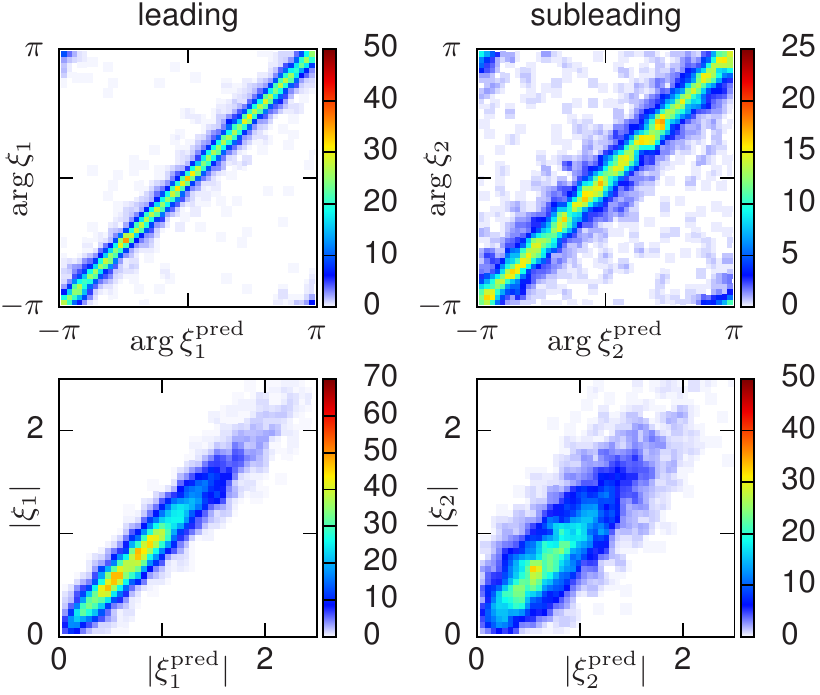}
		\caption{Angle and magnitude correlations between the flow and and the two term predictor in central collisions.\label{phi}}
\end{figure*}

In \Fig{phi} we show the correlations between the flow and its predictor for
both the angles and magnitudes. The subleading flow direction correlates  well
with the predictor, and there is reasonable correlation for the magnitude as
well.

\pagebreak

\bibliography{master}

\begin{thebibliography}{20}%
\makeatletter
\providecommand \@ifxundefined [1]{%
 \@ifx{#1\undefined}
}%
\providecommand \@ifnum [1]{%
 \ifnum #1\expandafter \@firstoftwo
 \else \expandafter \@secondoftwo
 \fi
}%
\providecommand \@ifx [1]{%
 \ifx #1\expandafter \@firstoftwo
 \else \expandafter \@secondoftwo
 \fi
}%
\providecommand \natexlab [1]{#1}%
\providecommand \enquote  [1]{``#1''}%
\providecommand \bibnamefont  [1]{#1}%
\providecommand \bibfnamefont [1]{#1}%
\providecommand \citenamefont [1]{#1}%
\providecommand \href@noop [0]{\@secondoftwo}%
\providecommand \href [0]{\begingroup \@sanitize@url \@href}%
\providecommand \@href[1]{\@@startlink{#1}\@@href}%
\providecommand \@@href[1]{\endgroup#1\@@endlink}%
\providecommand \@sanitize@url [0]{\catcode `\\12\catcode `\$12\catcode
  `\&12\catcode `\#12\catcode `\^12\catcode `\_12\catcode `\%12\relax}%
\providecommand \@@startlink[1]{}%
\providecommand \@@endlink[0]{}%
\providecommand \url  [0]{\begingroup\@sanitize@url \@url }%
\providecommand \@url [1]{\endgroup\@href {#1}{\urlprefix }}%
\providecommand \urlprefix  [0]{URL }%
\providecommand \Eprint [0]{\href }%
\providecommand \doibase [0]{http://dx.doi.org/}%
\providecommand \selectlanguage [0]{\@gobble}%
\providecommand \bibinfo  [0]{\@secondoftwo}%
\providecommand \bibfield  [0]{\@secondoftwo}%
\providecommand \translation [1]{[#1]}%
\providecommand \BibitemOpen [0]{}%
\providecommand \bibitemStop [0]{}%
\providecommand \bibitemNoStop [0]{.\EOS\space}%
\providecommand \EOS [0]{\spacefactor3000\relax}%
\providecommand \BibitemShut  [1]{\csname bibitem#1\endcsname}%
\let\auto@bib@innerbib\@empty
\bibitem [{\citenamefont {Heinz}\ and\ \citenamefont
  {Snellings}(2013)}]{Heinz:2013th}%
  \BibitemOpen
  \bibfield  {author} {\bibinfo {author} {\bibfnamefont {Ulrich}\ \bibnamefont
  {Heinz}}\ and\ \bibinfo {author} {\bibfnamefont {Raimond}\ \bibnamefont
  {Snellings}},\ }\bibfield  {title} {\enquote {\bibinfo {title} {{Collective
  flow and viscosity in relativistic heavy-ion collisions}},}\ }\href {\doibase
  10.1146/annurev-nucl-102212-170540} {\bibfield  {journal} {\bibinfo
  {journal} {Ann. Rev. Nucl. Part. Sci.}\ }\textbf {\bibinfo {volume} {63}},\
  \bibinfo {pages} {123--151} (\bibinfo {year} {2013})},\ \Eprint
  {http://arxiv.org/abs/1301.2826} {arXiv:1301.2826 [nucl-th]} \BibitemShut
  {NoStop}%
\bibitem [{\citenamefont {Luzum}\ and\ \citenamefont
  {Petersen}(2014)}]{Luzum:2013yya}%
  \BibitemOpen
  \bibfield  {author} {\bibinfo {author} {\bibfnamefont {Matthew}\ \bibnamefont
  {Luzum}}\ and\ \bibinfo {author} {\bibfnamefont {Hannah}\ \bibnamefont
  {Petersen}},\ }\bibfield  {title} {\enquote {\bibinfo {title} {{Initial State
  Fluctuations and Final State Correlations in Relativistic Heavy-Ion
  Collisions}},}\ }\href {\doibase 10.1088/0954-3899/41/6/063102} {\bibfield
  {journal} {\bibinfo  {journal} {J. Phys.}\ }\textbf {\bibinfo {volume}
  {G41}},\ \bibinfo {pages} {063102} (\bibinfo {year} {2014})},\ \Eprint
  {http://arxiv.org/abs/1312.5503} {arXiv:1312.5503 [nucl-th]} \BibitemShut
  {NoStop}%
\bibitem [{\citenamefont {Alver}\ and\ \citenamefont
  {Roland}(2010)}]{Alver:2010gr}%
  \BibitemOpen
  \bibfield  {author} {\bibinfo {author} {\bibfnamefont {B.}~\bibnamefont
  {Alver}}\ and\ \bibinfo {author} {\bibfnamefont {G.}~\bibnamefont {Roland}},\
  }\bibfield  {title} {\enquote {\bibinfo {title} {{Collision geometry
  fluctuations and triangular flow in heavy-ion collisions}},}\ }\href
  {\doibase 10.1103/PhysRevC.82.039903, 10.1103/PhysRevC.81.054905} {\bibfield
  {journal} {\bibinfo  {journal} {Phys. Rev.}\ }\textbf {\bibinfo {volume}
  {C81}},\ \bibinfo {pages} {054905} (\bibinfo {year} {2010})},\ \bibinfo
  {note} {[Erratum: Phys. Rev.C82,039903(2010)]},\ \Eprint
  {http://arxiv.org/abs/1003.0194} {arXiv:1003.0194 [nucl-th]} \BibitemShut
  {NoStop}%
\bibitem [{\citenamefont {Gardim}\ \emph {et~al.}(2013)\citenamefont {Gardim},
  \citenamefont {Grassi}, \citenamefont {Luzum},\ and\ \citenamefont
  {Ollitrault}}]{Gardim:2012im}%
  \BibitemOpen
  \bibfield  {author} {\bibinfo {author} {\bibfnamefont {Fernando~G.}\
  \bibnamefont {Gardim}}, \bibinfo {author} {\bibfnamefont {Frederique}\
  \bibnamefont {Grassi}}, \bibinfo {author} {\bibfnamefont {Matthew}\
  \bibnamefont {Luzum}}, \ and\ \bibinfo {author} {\bibfnamefont {Jean-Yves}\
  \bibnamefont {Ollitrault}},\ }\bibfield  {title} {\enquote {\bibinfo {title}
  {{Breaking of factorization of two-particle correlations in
  hydrodynamics}},}\ }\href {\doibase 10.1103/PhysRevC.87.031901} {\bibfield
  {journal} {\bibinfo  {journal} {Phys. Rev.}\ }\textbf {\bibinfo {volume}
  {C87}},\ \bibinfo {pages} {031901} (\bibinfo {year} {2013})},\ \Eprint
  {http://arxiv.org/abs/1211.0989} {arXiv:1211.0989 [nucl-th]} \BibitemShut
  {NoStop}%
\bibitem [{\citenamefont {Heinz}\ \emph {et~al.}(2013)\citenamefont {Heinz},
  \citenamefont {Qiu},\ and\ \citenamefont {Shen}}]{Heinz:2013bua}%
  \BibitemOpen
  \bibfield  {author} {\bibinfo {author} {\bibfnamefont {Ulrich}\ \bibnamefont
  {Heinz}}, \bibinfo {author} {\bibfnamefont {Zhi}\ \bibnamefont {Qiu}}, \ and\
  \bibinfo {author} {\bibfnamefont {Chun}\ \bibnamefont {Shen}},\ }\bibfield
  {title} {\enquote {\bibinfo {title} {{Fluctuating flow angles and anisotropic
  flow measurements}},}\ }\href {\doibase 10.1103/PhysRevC.87.034913}
  {\bibfield  {journal} {\bibinfo  {journal} {Phys. Rev.}\ }\textbf {\bibinfo
  {volume} {C87}},\ \bibinfo {pages} {034913} (\bibinfo {year} {2013})},\
  \Eprint {http://arxiv.org/abs/1302.3535} {arXiv:1302.3535 [nucl-th]}
  \BibitemShut {NoStop}%
\bibitem [{\citenamefont {Kozlov}\ \emph {et~al.}(2014)\citenamefont {Kozlov},
  \citenamefont {Luzum}, \citenamefont {Denicol}, \citenamefont {Jeon},\ and\
  \citenamefont {Gale}}]{Kozlov:2014fqa}%
  \BibitemOpen
  \bibfield  {author} {\bibinfo {author} {\bibfnamefont {Igor}\ \bibnamefont
  {Kozlov}}, \bibinfo {author} {\bibfnamefont {Matthew}\ \bibnamefont {Luzum}},
  \bibinfo {author} {\bibfnamefont {Gabriel}\ \bibnamefont {Denicol}}, \bibinfo
  {author} {\bibfnamefont {Sangyong}\ \bibnamefont {Jeon}}, \ and\ \bibinfo
  {author} {\bibfnamefont {Charles}\ \bibnamefont {Gale}},\ }\bibfield  {title}
  {\enquote {\bibinfo {title} {{Transverse momentum structure of pair
  correlations as a signature of collective behavior in small collision
  systems}},}\ }\href@noop {} {\  (\bibinfo {year} {2014})},\ \Eprint
  {http://arxiv.org/abs/1405.3976} {arXiv:1405.3976 [nucl-th]} \BibitemShut
  {NoStop}%
\bibitem [{\citenamefont {Bhalerao}\ \emph {et~al.}(2015)\citenamefont
  {Bhalerao}, \citenamefont {Ollitrault}, \citenamefont {Pal},\ and\
  \citenamefont {Teaney}}]{Bhalerao:2014mua}%
  \BibitemOpen
  \bibfield  {author} {\bibinfo {author} {\bibfnamefont {Rajeev~S.}\
  \bibnamefont {Bhalerao}}, \bibinfo {author} {\bibfnamefont {Jean-Yves}\
  \bibnamefont {Ollitrault}}, \bibinfo {author} {\bibfnamefont {Subrata}\
  \bibnamefont {Pal}}, \ and\ \bibinfo {author} {\bibfnamefont {Derek}\
  \bibnamefont {Teaney}},\ }\bibfield  {title} {\enquote {\bibinfo {title}
  {{Principal component analysis of event-by-event fluctuations}},}\ }\href
  {\doibase 10.1103/PhysRevLett.114.152301} {\bibfield  {journal} {\bibinfo
  {journal} {Phys. Rev. Lett.}\ }\textbf {\bibinfo {volume} {114}},\ \bibinfo
  {pages} {152301} (\bibinfo {year} {2015})},\ \Eprint
  {http://arxiv.org/abs/1410.7739} {arXiv:1410.7739 [nucl-th]} \BibitemShut
  {NoStop}%
\bibitem [{\citenamefont {Teaney}\ and\ \citenamefont
  {Yan}(2011)}]{Teaney:2010vd}%
  \BibitemOpen
  \bibfield  {author} {\bibinfo {author} {\bibfnamefont {Derek}\ \bibnamefont
  {Teaney}}\ and\ \bibinfo {author} {\bibfnamefont {Li}~\bibnamefont {Yan}},\
  }\bibfield  {title} {\enquote {\bibinfo {title} {{Triangularity and Dipole
  Asymmetry in Heavy Ion Collisions}},}\ }\href {\doibase
  10.1103/PhysRevC.83.064904} {\bibfield  {journal} {\bibinfo  {journal} {Phys.
  Rev.}\ }\textbf {\bibinfo {volume} {C83}},\ \bibinfo {pages} {064904}
  (\bibinfo {year} {2011})},\ \Eprint {http://arxiv.org/abs/1010.1876}
  {arXiv:1010.1876 [nucl-th]} \BibitemShut {NoStop}%
\bibitem [{\citenamefont {Qiu}\ and\ \citenamefont {Heinz}(2011)}]{Qiu:2011iv}%
  \BibitemOpen
  \bibfield  {author} {\bibinfo {author} {\bibfnamefont {Zhi}\ \bibnamefont
  {Qiu}}\ and\ \bibinfo {author} {\bibfnamefont {Ulrich~W.}\ \bibnamefont
  {Heinz}},\ }\bibfield  {title} {\enquote {\bibinfo {title} {{Event-by-event
  shape and flow fluctuations of relativistic heavy-ion collision
  fireballs}},}\ }\href {\doibase 10.1103/PhysRevC.84.024911} {\bibfield
  {journal} {\bibinfo  {journal} {Phys. Rev.}\ }\textbf {\bibinfo {volume}
  {C84}},\ \bibinfo {pages} {024911} (\bibinfo {year} {2011})},\ \Eprint
  {http://arxiv.org/abs/1104.0650} {arXiv:1104.0650 [nucl-th]} \BibitemShut
  {NoStop}%
\bibitem [{\citenamefont {Baier}\ \emph {et~al.}(2008)\citenamefont {Baier},
  \citenamefont {Romatschke}, \citenamefont {Son}, \citenamefont {Starinets},\
  and\ \citenamefont {Stephanov}}]{Baier:2007ix}%
  \BibitemOpen
  \bibfield  {author} {\bibinfo {author} {\bibfnamefont {Rudolf}\ \bibnamefont
  {Baier}}, \bibinfo {author} {\bibfnamefont {Paul}\ \bibnamefont
  {Romatschke}}, \bibinfo {author} {\bibfnamefont {Dam~Thanh}\ \bibnamefont
  {Son}}, \bibinfo {author} {\bibfnamefont {Andrei~O.}\ \bibnamefont
  {Starinets}}, \ and\ \bibinfo {author} {\bibfnamefont {Mikhail~A.}\
  \bibnamefont {Stephanov}},\ }\bibfield  {title} {\enquote {\bibinfo {title}
  {{Relativistic viscous hydrodynamics, conformal invariance, and
  holography}},}\ }\href {\doibase 10.1088/1126-6708/2008/04/100} {\bibfield
  {journal} {\bibinfo  {journal} {JHEP}\ }\textbf {\bibinfo {volume} {04}},\
  \bibinfo {pages} {100} (\bibinfo {year} {2008})},\ \Eprint
  {http://arxiv.org/abs/0712.2451} {arXiv:0712.2451 [hep-th]} \BibitemShut
  {NoStop}%
\bibitem [{\citenamefont {Dusling}\ and\ \citenamefont
  {Teaney}(2008)}]{Dusling:2007gi}%
  \BibitemOpen
  \bibfield  {author} {\bibinfo {author} {\bibfnamefont {K.}~\bibnamefont
  {Dusling}}\ and\ \bibinfo {author} {\bibfnamefont {D.}~\bibnamefont
  {Teaney}},\ }\bibfield  {title} {\enquote {\bibinfo {title} {{Simulating
  elliptic flow with viscous hydrodynamics}},}\ }\href {\doibase
  10.1103/PhysRevC.77.034905} {\bibfield  {journal} {\bibinfo  {journal} {Phys.
  Rev.}\ }\textbf {\bibinfo {volume} {C77}},\ \bibinfo {pages} {034905}
  (\bibinfo {year} {2008})},\ \Eprint {http://arxiv.org/abs/0710.5932}
  {arXiv:0710.5932 [nucl-th]} \BibitemShut {NoStop}%
\bibitem [{\citenamefont {Teaney}\ and\ \citenamefont
  {Yan}(2012)}]{Teaney:2012ke}%
  \BibitemOpen
  \bibfield  {author} {\bibinfo {author} {\bibfnamefont {Derek}\ \bibnamefont
  {Teaney}}\ and\ \bibinfo {author} {\bibfnamefont {Li}~\bibnamefont {Yan}},\
  }\bibfield  {title} {\enquote {\bibinfo {title} {{Non linearities in the
  harmonic spectrum of heavy ion collisions with ideal and viscous
  hydrodynamics}},}\ }\href {\doibase 10.1103/PhysRevC.86.044908} {\bibfield
  {journal} {\bibinfo  {journal} {Phys. Rev.}\ }\textbf {\bibinfo {volume}
  {C86}},\ \bibinfo {pages} {044908} (\bibinfo {year} {2012})},\ \Eprint
  {http://arxiv.org/abs/1206.1905} {arXiv:1206.1905 [nucl-th]} \BibitemShut
  {NoStop}%
\bibitem [{\citenamefont {Alver}\ \emph {et~al.}(2008)\citenamefont {Alver},
  \citenamefont {Baker}, \citenamefont {Loizides},\ and\ \citenamefont
  {Steinberg}}]{Alver:2008aq}%
  \BibitemOpen
  \bibfield  {author} {\bibinfo {author} {\bibfnamefont {B.}~\bibnamefont
  {Alver}}, \bibinfo {author} {\bibfnamefont {M.}~\bibnamefont {Baker}},
  \bibinfo {author} {\bibfnamefont {C.}~\bibnamefont {Loizides}}, \ and\
  \bibinfo {author} {\bibfnamefont {P.}~\bibnamefont {Steinberg}},\ }\bibfield
  {title} {\enquote {\bibinfo {title} {{The PHOBOS Glauber Monte Carlo}},}\
  }\href@noop {} {\  (\bibinfo {year} {2008})},\ \Eprint
  {http://arxiv.org/abs/0805.4411} {arXiv:0805.4411 [nucl-ex]} \BibitemShut
  {NoStop}%
\bibitem [{\citenamefont {Qiu}(2013)}]{Qiu:2013wca}%
  \BibitemOpen
  \bibfield  {author} {\bibinfo {author} {\bibfnamefont {Zhi}\ \bibnamefont
  {Qiu}},\ }\emph {\bibinfo {title} {{Event-by-event Hydrodynamic Simulations
  for Relativistic Heavy-ion Collisions}}},\ \href
  {https://inspirehep.net/record/1247294/files/arXiv:1308.2182.pdf} {Ph.D.
  thesis},\ \bibinfo  {school} {Ohio State U.} (\bibinfo {year} {2013}),\
  \Eprint {http://arxiv.org/abs/1308.2182} {arXiv:1308.2182 [nucl-th]}
  \BibitemShut {NoStop}%
\bibitem [{\citenamefont {Laine}\ and\ \citenamefont
  {Schroder}(2006)}]{Laine:2006cp}%
  \BibitemOpen
  \bibfield  {author} {\bibinfo {author} {\bibfnamefont {Mikko}\ \bibnamefont
  {Laine}}\ and\ \bibinfo {author} {\bibfnamefont {York}\ \bibnamefont
  {Schroder}},\ }\bibfield  {title} {\enquote {\bibinfo {title} {{Quark mass
  thresholds in QCD thermodynamics}},}\ }\href {\doibase
  10.1103/PhysRevD.73.085009} {\bibfield  {journal} {\bibinfo  {journal} {Phys.
  Rev.}\ }\textbf {\bibinfo {volume} {D73}},\ \bibinfo {pages} {085009}
  (\bibinfo {year} {2006})},\ \Eprint {http://arxiv.org/abs/hep-ph/0603048}
  {arXiv:hep-ph/0603048 [hep-ph]} \BibitemShut {NoStop}%
\bibitem [{\citenamefont {Luzum}\ and\ \citenamefont
  {Romatschke}(2008)}]{Luzum:2008cw}%
  \BibitemOpen
  \bibfield  {author} {\bibinfo {author} {\bibfnamefont {Matthew}\ \bibnamefont
  {Luzum}}\ and\ \bibinfo {author} {\bibfnamefont {Paul}\ \bibnamefont
  {Romatschke}},\ }\bibfield  {title} {\enquote {\bibinfo {title} {{Conformal
  Relativistic Viscous Hydrodynamics: Applications to RHIC results at
  s(NN)**(1/2) = 200-GeV}},}\ }\href {\doibase 10.1103/PhysRevC.78.034915,
  10.1103/PhysRevC.79.039903} {\bibfield  {journal} {\bibinfo  {journal} {Phys.
  Rev.}\ }\textbf {\bibinfo {volume} {C78}},\ \bibinfo {pages} {034915}
  (\bibinfo {year} {2008})},\ \bibinfo {note} {[Erratum: Phys.
  Rev.C79,039903(2009)]},\ \Eprint {http://arxiv.org/abs/0804.4015}
  {arXiv:0804.4015 [nucl-th]} \BibitemShut {NoStop}%
\bibitem [{\citenamefont {Başar}\ and\ \citenamefont
  {Teaney}(2014)}]{Basar:2013hea}%
  \BibitemOpen
  \bibfield  {author} {\bibinfo {author} {\bibfnamefont {Gökçe}\ \bibnamefont
  {Başar}}\ and\ \bibinfo {author} {\bibfnamefont {Derek}\ \bibnamefont
  {Teaney}},\ }\bibfield  {title} {\enquote {\bibinfo {title} {{Scaling
  relation between pA and AA collisions}},}\ }\href {\doibase
  10.1103/PhysRevC.90.054903} {\bibfield  {journal} {\bibinfo  {journal} {Phys.
  Rev.}\ }\textbf {\bibinfo {volume} {C90}},\ \bibinfo {pages} {054903}
  (\bibinfo {year} {2014})},\ \Eprint {http://arxiv.org/abs/1312.6770}
  {arXiv:1312.6770 [nucl-th]} \BibitemShut {NoStop}%
\bibitem [{\citenamefont {Gardim}\ \emph {et~al.}(2012)\citenamefont {Gardim},
  \citenamefont {Grassi}, \citenamefont {Luzum},\ and\ \citenamefont
  {Ollitrault}}]{Gardim:2011xv}%
  \BibitemOpen
  \bibfield  {author} {\bibinfo {author} {\bibfnamefont {Fernando~G.}\
  \bibnamefont {Gardim}}, \bibinfo {author} {\bibfnamefont {Frederique}\
  \bibnamefont {Grassi}}, \bibinfo {author} {\bibfnamefont {Matthew}\
  \bibnamefont {Luzum}}, \ and\ \bibinfo {author} {\bibfnamefont {Jean-Yves}\
  \bibnamefont {Ollitrault}},\ }\bibfield  {title} {\enquote {\bibinfo {title}
  {{Mapping the hydrodynamic response to the initial geometry in heavy-ion
  collisions}},}\ }\href {\doibase 10.1103/PhysRevC.85.024908} {\bibfield
  {journal} {\bibinfo  {journal} {Phys. Rev.}\ }\textbf {\bibinfo {volume}
  {C85}},\ \bibinfo {pages} {024908} (\bibinfo {year} {2012})},\ \Eprint
  {http://arxiv.org/abs/1111.6538} {arXiv:1111.6538 [nucl-th]} \BibitemShut
  {NoStop}%
\bibitem [{\citenamefont {Floerchinger}\ and\ \citenamefont
  {Wiedemann}(2014)}]{Floerchinger:2014fta}%
  \BibitemOpen
  \bibfield  {author} {\bibinfo {author} {\bibfnamefont {Stefan}\ \bibnamefont
  {Floerchinger}}\ and\ \bibinfo {author} {\bibfnamefont {Urs~Achim}\
  \bibnamefont {Wiedemann}},\ }\bibfield  {title} {\enquote {\bibinfo {title}
  {{Statistics of initial density perturbations in heavy ion collisions and
  their fluid dynamic response}},}\ }\href {\doibase 10.1007/JHEP08(2014)005}
  {\bibfield  {journal} {\bibinfo  {journal} {JHEP}\ }\textbf {\bibinfo
  {volume} {08}},\ \bibinfo {pages} {005} (\bibinfo {year} {2014})},\ \Eprint
  {http://arxiv.org/abs/1405.4393} {arXiv:1405.4393 [hep-ph]} \BibitemShut
  {NoStop}%
\bibitem [{\citenamefont {Gardim}\ \emph {et~al.}(2015)\citenamefont {Gardim},
  \citenamefont {Noronha-Hostler}, \citenamefont {Luzum},\ and\ \citenamefont
  {Grassi}}]{Gardim:2014tya}%
  \BibitemOpen
  \bibfield  {author} {\bibinfo {author} {\bibfnamefont {Fernando~G.}\
  \bibnamefont {Gardim}}, \bibinfo {author} {\bibfnamefont {Jacquelyn}\
  \bibnamefont {Noronha-Hostler}}, \bibinfo {author} {\bibfnamefont {Matthew}\
  \bibnamefont {Luzum}}, \ and\ \bibinfo {author} {\bibfnamefont
  {Frédérique}\ \bibnamefont {Grassi}},\ }\bibfield  {title} {\enquote
  {\bibinfo {title} {{Effects of viscosity on the mapping of initial to final
  state in heavy ion collisions}},}\ }\href {\doibase
  10.1103/PhysRevC.91.034902} {\bibfield  {journal} {\bibinfo  {journal} {Phys.
  Rev.}\ }\textbf {\bibinfo {volume} {C91}},\ \bibinfo {pages} {034902}
  (\bibinfo {year} {2015})},\ \Eprint {http://arxiv.org/abs/1411.2574}
  {arXiv:1411.2574 [nucl-th]} \BibitemShut {NoStop}%
\end{thebibliography}%

\end{document}